\documentstyle[12pt,pictex,epsfig,amssymb,amsbsy]{article}

\renewcommand{\thefootnote}{\fnsymbol{footnote}}
\newcommand{\newsection}{
\setcounter{equation}{0}
\section}

\def\appendix#1{
  \addtocounter{section}{1}
  \setcounter{equation}{0}
  \renewcommand{\thesection}{\Alph{section}}
  \section*{Appendix \thesection\protect\indent \parbox[t]{11.715cm} {#1}}
  \addcontentsline{toc}{section}{Appendix \thesection\ \ \ #1}
  }

\newcommand{\bbox}[1]{\boldsymbol{#1}}
\newcommand {\defeq}{\stackrel{\rm def}{=}}
\newcommand{\tr}[1]{\:{\rm tr}\,#1}

\def\e{{\,\rm e}\,}

\def\d{{\rm d}}
\def\r{{r}}
\def\i{{\rm i}}

\newcommand{\rf}[1]{(\ref{#1})}
\newcommand{\eq}[1]{Eq.~(\ref{#1})}
\def\be{\begin{equation}}
\def\ee{\end{equation}}
\def\beq{\begin{equation}}
\def\eeq{\end{equation}}
\def\bea{\begin{eqnarray}}
\def\eea{\end{eqnarray}}
\def\LA{\left\langle}
\def\RA{\right\rangle}

\newcommand{\non}{\nonumber \\*}

\newcommand{\ra}{\rightarrow}
\begin{document}

\begin{titlepage}
\begin{flushright}
ITEP--TH--54/02\\
hep-th/0210256\\
October, 2002
\end{flushright}
\vspace{1.3cm}

\begin{center}
{\LARGE Light-Cone Wilson Loops 
and \\[.4cm] the String/Gauge Correspondence}\\
\vspace{1.4cm}
{\large Yuri Makeenko}\footnote{E--mail:
makeenko@itep.ru \ \ \ \
 makeenko@nbi.dk \ } \\
\vskip 0.2 cm
{\it Institute of Theoretical and Experimental Physics,}
\\ {\it B. Cheremushkinskaya 25, 117259 Moscow, Russia}
\\ \vskip .1 cm
and  \\  \vskip .1 cm
{\it The Niels Bohr Institute,} \\
{\it Blegdamsvej 17, 2100 Copenhagen {\O}, Denmark}
\end{center}
\vskip 1.5 cm
\begin{abstract}
We investigate a $\Pi$-shape Wilson loop in ${\cal N}=4$ super
Yang--Mills theory, which lies partially at the light-cone, and consider 
an associated open superstring in \mbox{$AdS_5 \times S^5$}.
We discuss how this Wilson loop determines the anomalous dimensions of 
conformal operators with large Lorentz spin and present an explicit 
calculation in perturbation theory to order $\lambda$.
We find the minimal surface in the supergravity approximation, 
that reproduces 
the Gubser, Klebanov and Polyakov prediction for the anomalous dimensions 
at large $\lambda=g_{\rm YM}^2 N$, 
and discuss its quantum-mechanical interpretation.
\end{abstract}

\end{titlepage}
\setcounter{page}{2}
\renewcommand{\thefootnote}{\arabic{footnote}}
\setcounter{footnote}{0}

\newsection{Introduction}

There is a long-standing belief%
\footnote{See e.g.\ Ref.~\cite{Mak02} for an introduction and 
review of the old works on the string/gauge correspondence.} 
that $SU(N)$ Yang--Mills theory
is equivalent at large $N$ to a free string, while the $1/N$-expansion  
corresponds to interactions of the string.
A great recent progress along this line 
is associated for ${\cal N}=4$ super Yang--Mills theory (SYM) with 
the AdS/CFT correspondence~\cite{Mal98a}, 
where the strong-coupling limit
of SYM is described by supergravity in anti-de Sitter space 
$AdS_5 \times S^5$. 

Among the most interesting predictions of the AdS/CFT correspondence
for the strong-coupling limit of SYM, let us mention
the calculation~\cite{GKP98,Wit98}
of anomalous dimensions of certain operators 
(see Ref.~\cite{AGMOO99} for a review) and that~\cite{Mal98b,RY98} of the 
Euclidean-space rectangular Wilson loop determining the interaction
potential. The former is given by the spectrum
of excitations in $AdS$ space, while the latter is given by
the minimal surface formed by the worldsheet of an open string whose ends lie
at the loop in the boundary of $AdS_5 \times S^5$.
The computations of the Wilson loops in the supergravity approximation
were performed
also for circular loops~\cite{BCFM98,DGO99} and loops with cusps~\cite{DGO99}.
The circular Wilson loop has then been exactly calculated~\cite{ESZ00}
in SYM to all orders in the 't~Hooft coupling $\lambda=g_{\rm YM}^2 N$.   
The result provided not only a beautiful test of the AdS/CFT correspondence
at large $\lambda$ but also a challenging prediction for IIB superstring
in the $AdS_5 \times S^5$ background~\cite{DG00}.

Yet another remarkable test of the string/gauge correspondence 
concerns~\cite{BMN02} a
certain class of operators in SYM, whose anomalous dimensions can be 
exactly computed as a function of $\lambda$ both in string theory
and under some mild assumptions in SYM~\cite{GMR02,Z}.
The exact computation in string theory is possible because 
the anomalous dimensions of these
BMN operators correspond to the spectrum of states
with large angular momentum associated with rotation of an infinitely 
short closed string around the equator of $S^5$.

Rotating similarly a long closed folded string in $AdS_5$, a very
interesting prediction concerning the strong-coupling limit
of the anomalous dimensions of twist (= bare dimension minus Lorentz spin $n$)
two operators has been obtained recently in Ref.~\cite{GKP02}:
\be
\Delta -n = f(\lambda) \ln {n} 
\label{GKP}
\ee
for large $n$,
where
\be
f(\lambda)=\frac{\sqrt{\lambda}}{\pi}
+{\cal O}\left((\sqrt{\lambda})^0\right)
\label{GKP1}
\ee
and $\lambda=g_{\rm YM}^2 N$ -- the 't~Hooft coupling in SYM -- is large.
The correction ${\cal O}((\sqrt{\lambda})^0)$ 
has been calculated~\cite{FT02}
and it has been shown~\cite{Kru02} very recently how 
the GKP result can be reproduced via a minimal surface
of an open string spanned in the boundary by the loop with a cusp.

Equations~\rf{GKP} and \rf{GKP1} 
were derived~\cite{GKP02} ignoring the $S^5$ part of $AdS_5 \times S^5$,
which is responsible for supersymmetry, and possess the features expected
for the anomalous dimension in ordinary (nonsupersymmetric) 
Yang--Mills theory.
There are arguments for this result to be valid in ordinary Yang--Mills theory
as well. This would lead us to very interesting predictions 
for the strong-coupling limit of QCD!

The goal of this paper is to further study the string/gauge correspondence
considering a Wilson loop with cusps, which
partially lies at the light cone, 
 on the SYM side and minimal surfaces in $AdS$ space
associated
with an open string ending at the loop in the boundary. 
We extend the results of Ref.~\cite{KM93} to the SYM case and
show how the vacuum expectation value of this Wilson loop determines
the anomalous dimensions of the twist-two operators in SYM perturbation
theory. We find an appropriate minimal surface in $AdS$ space and show 
that it reproduces the GKP result~\rf{GKP}, \rf{GKP1}. 

This paper is organized as follows. After a brief excursion to QCD,
we define in Sect.~2 the light-cone Wilson loop in SYM, discuss its 
properties and demonstrate 
in perturbation theory how it gives the anomalous dimensions
of the twist-two operators. An explicit calculation 
is presented to order $\lambda$. In Sect.~3 we find a solution for the
minimal surface associated with the worldsheet of 
an open string in the AdS background,
the ends of which lie at the loop in the boundary, and demonstrate
how it reproduces Eqs.~\rf{GKP} and \rf{GKP1}. We discuss some unusual
properties of this solution and its quantum-mechanical interpretation
as tunneling in an analogous mechanical problem.

\newsection{Light-cone Wilson loop in SYM}

\subsection{The set up}

The relation between the anomalous dimensions of hadronic operators 
${\cal O}_n$ with the Lorentz spin $n$
in QCD and the renormalization of open Wilson loops with quarks at the ends 
is well-known~\cite{CD81,BB88}. 
A crucial role in this correspondence
is played~\cite{Bro80,Mak81,Ond82} 
by conformal operators which are multiplicatively 
renormalizable at one loop. 

\begin{figure}
\vspace*{3mm}
\centering{
\input{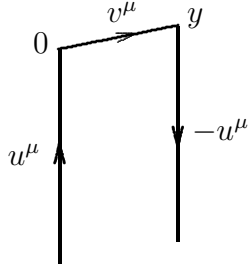}
}
\caption[Pi-shape Wilson loop]   
{$\Pi$-shape Wilson loop. The segment $[0,y^\mu\!=\!v^\mu T]$ 
lies at the light cone.
The loop is analytically given by \eq{parametrized}.}
   \label{fi:pishape}
\end{figure}

A convenient formulation of this approach was proposed by Korchemsky
and Marchesini~\cite{KM93} who considered a $\Pi$-shape Wilson loop 
\be
U(\Pi)= {\bbox P} \, \e^{\i \int_\Pi \d x^\mu A_\mu} 
\label{Pi-shape}
\ee
with the ends at infinity, whose middle segment lies at the light-cone.
It is depicted in Fig.~\ref{fi:pishape}.

This Wilson loop can be parametrized by
\be
x^\mu(t)=\left\{
\begin{array}{ll}\smallskip
u^\mu t & {\rm for}~~-\infty <t<0 \\ \smallskip
v^\mu t & {\rm for}~~0\leq t \leq T \\
v^\mu T -u^\mu (t-T) & {\rm for}~~T<t <\infty\,, \\
\end{array}
\right.
\label{parametrized}
\ee
where the unit vector $u^\mu$ is time-like ($u^2=-1$)
and the segment $[0,y^\mu\!=\!v^\mu T]$ lies at the light cone ($v^2=0$).
Without loss of generality we can choose 
\be
u^\mu=(-1,0,0,0)\,,~~~~~~~~ v^\mu=(1,1,0,0)
\label{choice}
\ee 
so that $y^\mu=(T,T,0,0)$. For the general case, we
define 
\be
L=u v \,T \,.
\label{defL}
\ee

The vacuum expectation value of such a Wilson loop 
\be
W(\Pi) \defeq \LA \frac 1N \tr{U(\Pi)} \RA
\label{filoop}
\ee
is a function of the ratio 
\be
\rho= \frac{L}{\epsilon}\,,
\label{rho}
\ee
where $L$ is defined by \eq{defL} and
$\epsilon$ is an ultraviolet
cutoff. It is the only dimensionless parameter which is present.

Though this $\Pi$-shape Wilson loop is {\em not}\/ renormalizable owing to
additional light-cone divergences, its logarithmic derivative
\be
\Gamma (\Pi)\defeq - \epsilon \frac{\d}{\d\, \epsilon} \log W(\Pi)\,.
\label{defGamma}
\ee
is multiplicatively renormalizable when expressed via the renormalized
coupling constant, so that $1/\epsilon$ in \eq{rho} 
can be replaced for $\Gamma (\Pi)$ by the renormalization-group scale $\mu$.

After an analytic continuation to imaginary $\rho$,
the logarithmic derivative~\rf{defGamma} determines the anomalous dimensions 
$\gamma_n$
of the hadro\-nic operators ${\cal O}_n$ with large $n$ by the 
formula~\cite{KM93}
\be
\left.\Gamma(\Pi)\right|_{\rho =-i n } =\gamma_n\,.
\label{gnvsGn}
\ee
Owing to the renormalizability of $\Gamma(\Pi)$, $\gamma_n$ depends
on $n$ logarithmically
\be
\gamma_n= f(\lambda) \log n\,,
\label{deff}
\ee
where $f(\lambda)$ is known to a few lower orders of perturbation theory.

Equations~\rf{defGamma} and \rf{gnvsGn} can be understood considering
the operator
\be
{\cal O}(C_{y0})= \bar \psi(y) U(C_{y0}) \psi(0)
\ee
associated with the straight open Wilson loop $C_{y0}$ with the 
(scalar or spinor) matter field $\psi$ attached at the end points.
Then
\be
\LA \psi(+\infty,\vec 0){\cal O}(C_{y0}) \bar \psi(-\infty,\vec y) \RA
\propto W (\Pi)
\ee
in the limit of the large mass of the matter field.
In other words, the $\Pi$-shape Wilson loop is associated with
the trajectory of a heavy particle which is at rest at the spacial
point $\vec 0$ for $-\infty<t<0$, moves from $\vec 0$ to
$\vec y$ along the light cone during the time $y_0$
and then stays at rest at $\vec y$ for $y_0\leq t <\infty$. 
It is clear from the analysis of perturbation-theory diagrams
that the anomalous dimensions $\gamma_n$ with large $n$ are correctly
reproduced in the large-mass limit. 

\subsection{Extension to ${\cal N}=4$ SYM}

We shall now extend the results, reviewed in the previous subsection,
to the ${\cal N}=4$ SYM.

We use the known supersymmetric 
extension~\cite{Mal98b,RY98,DGO99} of the Wilson loops in Minkowski
space and define
\be
U(C)= {\bbox P} \, \e^{\i \int 
\d s \,(\dot x^\mu A_\mu+
 |\dot x|\theta^i \Phi_i )} ,
\label{Pi-shapeSYM}
\ee
where $\Phi_i$ are six scalar fields and $\theta^i$ is a unit vector 
in ${\Bbb R}^6$. We shall be interested in the case when $C$ is
the $\Pi$-shape loop, the parametrization of which is
defined in \eq{parametrized}.

The difference between the Minkowski-space Wilson loops~\rf{Pi-shapeSYM}
and their Euclidean-space counterparts, which have been extensively
studied in the literature, is that the latter has an extra factor
of $\i$ in front of $\Phi_i$ in the exponent.
The Minkowski-space Wilson loop~\rf{Pi-shapeSYM}
is a BPS state, in particular,
for an infinite straight line which lies inside the light cone,
i.e.\ it can always be made parallel to the temporal axis by a Lorentz boost.
For such loops
\be
\LA \frac 1N \tr U\left(|\right) \RA=1 
\label{=1}
\ee
in a full analogy with straight lines in Euclidean space.

We shall be interested in anomalous dimensions of conformal operators
built out of the fields in SYM which belong to the adjoint
representation of $SU(N)$. Consequently, the Wilson loops are to be taken
in the adjoint representation:
\be
{\rm tr}_{\rm A} U (C) = |\tr{} U (C)|^2-1 \,.
\ee
Owing to the large-$N$ factorization, we have
\be
W_{\rm A}(C)\defeq \LA \frac 1{N^2\!-\!1} {\rm tr}_{\rm A} U (C)\RA 
\stackrel{N\ra\infty}{=} W^2(C) \,,
\label{A2F}
\ee
where the right-hand side is defined by Eqs.~\rf{filoop} 
and \rf{Pi-shapeSYM} for the fundamental representation.
Thus, the anomalous dimensions of such adjoint operators are twice larger
in the large-$N$ limit
than those in the fundamental representation.

\subsection{Order $\lambda$ of perturbation theory\label{ss:olambda}}

The expectation value of the adjoint Wilson loop to the order
$\lambda$ is the same as in the $U(1)$-case and is given by
\bea
W_{\rm A}(\Pi)
&=& 1 -  \frac\lambda2 \int\limits_{-\infty}^{+\infty} \d t_1 
\int\limits _{-\infty}^{+\infty}\d t_2
\Big[\dot x^\mu(t_1)\dot x_\mu(t_2)  +|\dot x(t_1)| |\dot x(t_2)| \Big]
D\big( x(t_1)- x(t_2)  \big)\,, \non
&&
\label{Wl2}
\eea
where the factor of $1/2$ is related to 
the normalization of the $SU(N)$ generators $t^a$. 

The scalar propagator in $d$ dimensions reads as
\be
D(x)=\frac{\Gamma(d/2-1)}{4\pi^{d/2}} (x^2)^{1-d/2} \,.
\ee
The relative sign of the two
terms in square brackets in \eq{Wl2} is minus for the temporal component
since the
gauge-field propagator in Minkowski space is
\be
D_{\mu\nu}(x)= \eta_{\mu\nu}D(x)\,,~~~~~~\eta_{\mu\nu}={\rm diag}\;
(-+++)
\ee
in the Feynman gauge.

To regularize divergent integrals, we shall use either
dimensional regularization when $0<4-d\ll 1$ or a smearing when
$D(x)$ is substituted by
\be
D_\epsilon(x)=\frac{1}{4\pi^2 (x^2+\epsilon^2)}\,.
\label{smeared}
\ee
Though this smearing is not gauge invariant, it will be enough for our 
purposes since the ${\cal N}=4$ SYM has no charge renormalization
in $d=4$ dimensions, so the only role of the regularization is to
regularize the light-cone or cusp singularities of the 
expectation value of the Wilson loop.

\begin{figure}
\vspace*{3mm}
\centering{
\input{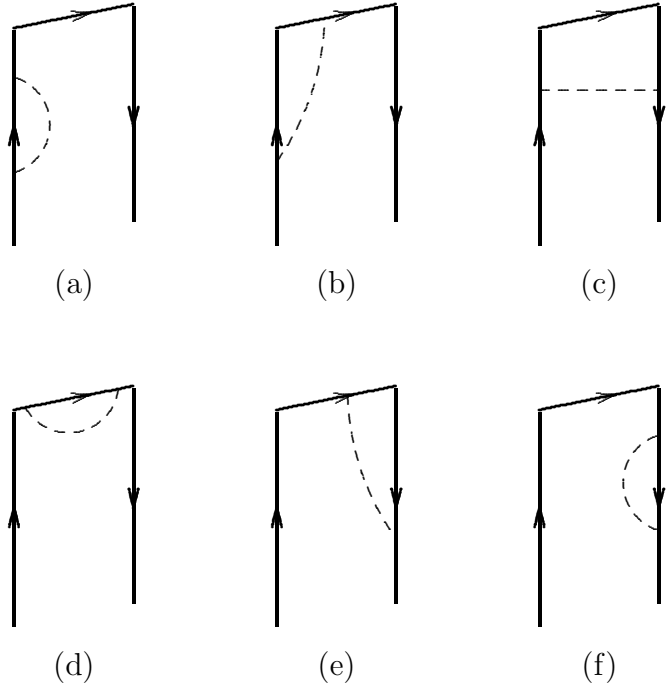}
}
\caption[Six diagrams]   
{Diagrams of the order $\lambda$ for the expectation value of the 
$\Pi$-shape Wilson loop. The dashed lines represent either scalar
or gauge-field propagators. Only the diagrams in Figs.~(b) and (e)
contribute to the anomalous dimension $\gamma_n$.}
   \label{fi:orderl}
\end{figure}

Depending on which segment of the loop the points  $x(t_1)$ and $x(t_2)$
belong to, the right-hand side of \eq{Wl2} can be represented as the 
sum of the six diagrams in Fig.~\ref{fi:orderl}.
The diagrams in Figs.~\ref{fi:orderl}(a) and (f) vanish quite similarly
to the case of an infinite straight line (cf.\ \eq{=1}),
since this straight segments are time-like. 
The diagram in Fig.~\ref{fi:orderl}(d) vanishes since the ends of 
the propagators are at the light-cone where $|\dot x|=0$.
For the same reason
the contribution of the diagrams in Figs.~\ref{fi:orderl}(b) and (e) 
is as in Ref.~\cite{KM93}: these of the scalars vanish. 

The diagram in Fig.~\ref{fi:orderl}(b) reads 
for the regularization~\rf{smeared} as
\be
W^{(1)}_{{\rm (b)}}=\frac{\lambda}{4\pi^2} uv
\int\limits_0^\infty \d s \int\limits_0^T \d t 
\frac1{s^2 -2uv st-\epsilon^2}\, ,
\label{dia(b)}
\ee
where the light-cone segment is parametrized by $t$ and the
infinite segment is parametrized by $s$.  
Using the variable \rf{defL},
we rewrite \eq{dia(b)} as
\bea
W^{(1)}_{{\rm (b)}}&=& \frac{\lambda}{4\pi^2}
\int \limits_0^\infty \d s \int\limits_0^L \d x 
\frac1{s^2 -2 sx-\epsilon^2} \,, \non
&=&- \frac{\lambda}{8\pi^2}
\int\limits_0^L \d x 
\frac1{\sqrt{x^2+\epsilon^2}} 
\log {\frac{x+\sqrt{x^2+\epsilon^2}}{x-\sqrt{x^2+\epsilon^2}}} \,.
\label{bfiol}
\eea
Substituting 
\be
\frac{L}\epsilon=-\i n\,, 
\label{continuing}
\ee
as is prescribed by \eq{gnvsGn}, we get finally 
\bea
W^{(1)}_{{\rm (b)}}
&=& -\frac{\lambda}{8\pi^2} \left[
\frac 14 \log^2 \left( \frac{n+\sqrt{n^2-1}}{n-\sqrt{n^2-1}}  \right)  
+\frac{\pi^2}{4}
\right] \non
&\stackrel{n\rightarrow\infty}{=} & 
-\frac{\lambda}{8\pi^2} \left[ \log^2 (2n) + \frac{\pi^2}{4} \right].
\label{fiol}
\eea
The $\log^2$-term is as in Ref.~\cite{KM93}.

The diagram in Fig.~\ref{fi:orderl}(e) gives exactly the same result
as that  in Fig.~\ref{fi:orderl}(b).

The diagram in Fig.~\ref{fi:orderl}(c) connects the two infinite
vertical segments. It contributes to the interaction potential
(see e.g.\ Ref.~\cite{Mak02}, p.~255) but does not have the
$\log^2{(L/\epsilon)}$ which contributes to $\gamma_n$.
The diagram in Fig.~\ref{fi:orderl}(c) is therefore not essential for
the calculation of the anomalous dimension.

Adding the diagrams in Figs.~\ref{fi:orderl}(b) and (e), we obtain
for the anomalous dimension to order $\lambda$
\be
\gamma_n=\frac{\lambda}{2\pi^2} \log{n}
\label{smth}
\ee
which reproduces the result of
an explicit calculation in Ref.~\cite{KL00}.

It is worth noting that the right-hand side of \eq{fiol} became
real only after the analytic continuation~\rf{continuing}.
For real $L$ it has an imaginary part associated with the
divergence of the integral over $s$ in \eq{bfiol} for 
$s=x+\sqrt{x^2+\epsilon^2}$. The   
meaning of such an imaginary part
for the Wilson loops with cusps in Minkowski space is discussed in 
Ref.~\cite{KR87}.
This imaginary part will also appear in Sect.~\ref{minsurf}.

\subsection{Remark on cusp near the light cone}

\begin{figure}
\vspace*{3mm}
\centering{
\input{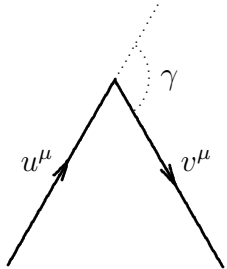}
}
\caption[Wilson loop with cusp]   
{$\Gamma$-shape Wilson loop having a cusp. 
The cusp angle $\gamma$ is given by \eq{cuspangle}.}
   \label{fi:cusp}
\end{figure}

As is pointed out in Ref.~\cite{KM93}, the anomalous dimension $\gamma_n$
is determined by the cusp anomalous dimension $\Gamma_{\rm cusp}$
associated with the renormalization of Wilson loops with cusps as depicted
in Fig.~\ref{fi:cusp}. 
The precise relation is as follows:
\be
\Gamma_{\rm cusp}\stackrel{\gamma\rightarrow\infty}{=} 
\frac{\gamma}{2} f(\lambda)\,,
\label{cuspGamma}
\ee
where $\gamma$ is the cusp angle defined in Minkowski space by
\be
\cosh{\gamma}=\frac {uv}{\sqrt{v^2}}
\label{cuspangle}
\ee
and the function $f(\lambda)$ enters \eq{deff}.

It is easy to obtain \eq{cuspGamma} taking the light-cone limit ($v^2\ra 0$)
of the renormaliza\-tion-group equation satisfied by~\rf{defGamma} and
noting that $\gamma\ra\infty$ in this limit.
The cusp anomalous dimension in the limit of large $\gamma$ was
discussed in Ref.~\cite{KR87}.

Note that the diagrams in Figs.~\ref{fi:orderl}(b) and (e) look
similar to the diagrams which give the cusp anomalous dimension.
The difference is that in our case one segment of the 
$\Gamma$-shape loop lies strictly at the light-cone, so the term
$\gamma \log{L/\epsilon}$ in the vacuum expectation value of the Wilson
loop with the cusp
is replaced by $(\log^2 {L/\epsilon})/2$ in our case.
We find it more convenient to deal directly with the light-cone
Wilson loop than to approach the light-cone first considering
the loop in Fig.~\ref{fi:cusp} at finite $\gamma$ and then taking 
the limit $\gamma\ra\infty$.

\newsection{Open string in $AdS_5\times S^5$ \label{minsurf}}

As is shown in Refs.~\cite{Mal98b,RY98}, the supersymmetric Wilson
loop~\rf{Pi-shapeSYM} is dual to an open string in $AdS_5\times S^5$,
the ends of which run along the contour $\{x^\mu(s), |x(s)|\theta^i\}$ at
the boundary of $AdS_5\times S^5$. In the supergravity limit, the string
worldsheet coincides with the minimal surface in $AdS_5$ bounded by
$x^\mu(s)$. This determines the asymptotic behavior of the Wilson loop
for large $\lambda$. 
While the original solutions, obtained in 
Refs.~\cite{Mal98b,RY98} for an (infinite) rectangle, 
in Refs.~\cite{BCFM98,DGO99} for a circle, and in Ref.~\cite{DGO99}
for the loop with a cusp depicted in Fig.~\ref{fi:cusp}, are
given in Euclidean space, their Minkowski-space analogs are
known~\cite{TZ02} in the former two cases.

\subsection{Finding the minimal surface}

\begin{figure}
\vspace*{3mm}
\centering{
\input{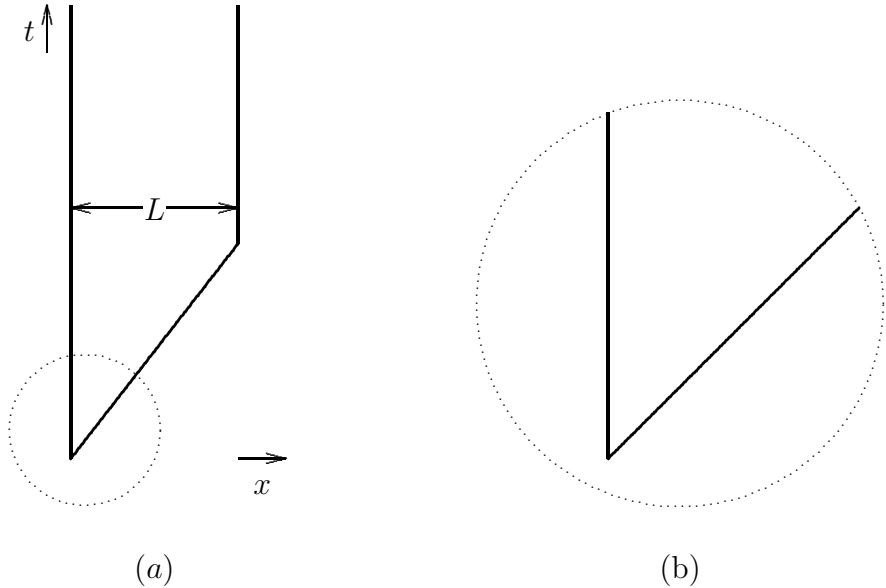}
}
\caption[Three diagrams for Wilson loop with cusp]   
{$\Pi$-shape loop (a) bounding the minimal surface.
The rotated segment lies at the light cone.
The $\log^2$-term comes from the region near the cusp which
is magnified in (b). There are two such regions associated with the
two cusps. The typical size of the magnified region is $\sim L$.}
   \label{magnify}
\end{figure}

To calculate the minimal surface for the Minkowski-space loop in 
Fig.~\ref{fi:pishape}, we parametrize the worldsheet by the coordinates
$t=x^0$ and $x=x^1$ for the choice~\rf{choice} of the $\Pi$-shape contour
as is depicted in Fig.~\ref{magnify}(a). 
The Nambu--Goto action of an open string  
in $AdS_3$ space with the metric given in the Poincar{\'e} coordinates by 
\be
\d s^2=R^2 \frac{-(\d x^0)^2+(\d x^1)^2+(\d z)^2}{z^2}
\label{AdSmetric}
\ee
reads as
\be
S=- \frac{R^2}{2\pi \alpha^\prime}
\int\limits_0^L \d x \int\limits_x^\infty \d t \frac{1}{z^2}
\sqrt{1+\left. z^\prime\right.^2 -\left. \dot z\right.^2}\,,
\label{NGaction}
\ee
where $\dot z = \d z/\d t$ and $z^\prime = \d z / \d x$.
We consider the $AdS_3$ subspace of $AdS_5$ space because we
set $x^2=x^3=0$ for the location of the minimal surface. 

The minimal surface $z(t,x)$ which is obtained by minimizing
the action~\rf{NGaction} depends in general both on $t$ and $x$.
However, the $\log^2$-term yielding the anomalous dimension comes
from the part of the minimal surface near the cusp as is shown
in Fig.~\ref{magnify}. In this case we substitute
\be
z(t,x)=\sqrt{t^2-x^2}\frac{1}{f\left(\frac xt\right)}
\label{ansatz}
\ee
so that only a function of the dimensionless ratio $x/t$ is to be
determined. This reminds the calculation of the cusp anomalous
dimension in Ref.~\cite{DGO99}.

Another important limiting case is when $t\gg x$, where $z$
does not depend on $t$ owing to translational symmetry of the problem.
This reproduces the calculation of the interaction potential
in Refs.~\cite{Mal98b,RY98} and perfectly agrees with what is said
at the end of Subsect.~\ref{ss:olambda} concerning
the diagram in Fig.~\ref{fi:orderl}(c) which contributes only to the
interaction potential rather than to the cusp anomalous dimension.

Before substituting the ansatz~\rf{ansatz} into \eq{NGaction},
it is convenient to parametrize the worldsheet
by the coordinates
\be
\r=x\,,~~~~~ \theta ={\rm arctanh}\,\frac{x}{t} 
\label{polar}
\ee
so that $0 <\r<L$ and $0 <\theta<\infty$. The two rays of the
contour in Fig.~\ref{magnify}(b) correspond to $\theta=0$ and
$\theta=\infty$. Then we rewrite the action~\rf{NGaction} as
\be
S=- 2\frac{R^2}{2\pi \alpha^\prime}\int \frac {\d \r}{\r} 
\int \d \theta \sqrt{f^4-f^2 +\left. f ^\prime \right.^2} \,,
\label{Srt}
\ee
where the factor of 2 is because the contour has two cusps.
The difference from its Euclidean counterpart of Ref.~\cite{DGO99}
is the sign in front of the $f^2$-term.

The function $f(\theta)$ obeys the boundary condition
\be
f(0)=\infty
\label{b.c.} 
\ee
for the minimal surface to end up in the boundary of $AdS$ space.
There is no boundary condition imposed on $f(\infty)$
since $z=0$ for the ansatz~\rf{ansatz} when $t=x$, i.e.\
 when the surface approaches the light cone in the boundary.
The light cone, given in $AdS$ space by $z^2=t^2-x^2$, corresponds to $f=1$, 
while the Poincar{\'e} horizon would be associated with $f=0$
in these coordinates.

In analogy with Refs.~\cite{Mal98b,DGO99}, the extremum of \rf{Srt}
 is given by the analytic function%
\footnote{More details are presented in Appendix~A.}
\be
\theta= \sqrt{2} \,{\rm arctanh}\, \sqrt{2(1-f^2)} 
-{\rm arctanh}\,\sqrt{1-f^2}-
\frac {\i \pi}{2} \left(\sqrt2 -1 \right).
\label{extremum}
\ee
This solution obeys~\rf{b.c.} for $\theta=0$ and 
$f\ra1/\sqrt{2}$ as $\theta\ra\infty$. 
When $f$ approaches $1/\sqrt{2}$, the minimal surface approaches
the one found in Ref.~\cite{Kru02} and given by 
\be
z=\sqrt{2(t^2-x^2)}\,.
\label{mikru}
\ee

Noting that $f(\theta)$, given by an inverse to \eq{extremum}, satisfies
\be
f^\prime = f \sqrt{1-f^2} \,(1-2f^2) \,, 
\ee
we see that the right-hand side of \eq{Srt} is linearly
divergent for $f\ra\infty$ and logarithmically divergent for 
$f\ra 1/\sqrt{2}$. After the standard subtraction of the linear divergence,
we have the function 
\bea
A(f)&=&\frac{R^2}{\pi \alpha^\prime} \int \frac{\d \r}{\r} \left\{
\int\limits_{f}^\infty \d f\left[ f\frac {\sqrt{f^2-1}}{f^2-1/2}-1 \right] -f
\right\}
\non &=&\i \frac{R^2}{\pi \alpha^\prime}\int \frac{\d \r}{\r} \Bigg\{
\frac{1}{2\sqrt{2}}\left[{\rm arctanh}\left(
\frac{\sqrt{1-f^2}}{\sqrt{2}-f}\right)+{\rm arctanh}\left(
\frac{\sqrt{1-f^2}}{\sqrt{2}+f}\right) \right] \non
 && \hspace*{3cm} - \sqrt{1-f^2} \Bigg\} 
\label{HJ}
\eea 
which gives the area of the part of the minimal surface from $\theta=0$
to $\theta=\theta(f)$.

For $\theta\ra\infty$ when $f\ra1/\sqrt{2}$, the first arctanh
in square
brackets logarithmically diverges. One more logarithmic
divergence comes from the integral of $\d \r/\r$.

To regularize these divergences, we proceed in the standard way~\cite{Mal98b}
shifting the boundary to $z=\epsilon$.
Then the consideration is quite similar to that in Ref.~\cite{DGO99}
for the case of the loop with a cusp.
From Eqs.~\rf{ansatz} and \rf{polar}, we have
\be
f(\theta)\sinh \theta < \frac{\r}{\epsilon}
\,.
\label{r>}
\ee
For $\theta\gg 1$ this implies,
\be
\theta < \theta_{\rm max}= \log{\frac{2\sqrt{2}\r}{\epsilon}} 
\label{thetamax}
\ee
and from \eq{extremum} we obtain
\be
f>f_{\rm min}=\frac{1}{\sqrt{2}} +\sqrt{2}\e^{-\sqrt{2} \theta_{\rm max}}\,.
\ee
Also $\r$ is bounded above by the value of the order of
$L$ which is the size of the magnified
region in Fig.~\ref{magnify}.

Evaluating the right-hand side of \eq{HJ} for $f=f_{\rm min}$,
we finally obtain the $\log^2$-term
\be
A(f_{\rm min})
=\i \frac{R^2}{2\pi \alpha^\prime} \int\frac{\d \r}{\r}
\, \theta_{\rm max}(\r)= \i \frac{R^2}{4\pi \alpha^\prime} 
\log^2{\frac{L}{\epsilon}} \,.
\label{Sfin}
\ee
It is easy to obtain the same answer directly form 
the first line in \eq{Srt}
substituting $f=1/\sqrt{2}$ for large $\theta$ which results 
in the term linear in $\theta_{\rm max}$. The latter procedure
is equivalent to saying that the area of the minimal surface is 
dominated by the contribution from its part described by \eq{mikru}
which was also the case in Ref.~\cite{Kru02}.
This is the reason why our results for the anomalous dimension
agree, while the solutions for the minimal surface are different. 

The appearance of the factor of $\i$ in \eq{Sfin} is remarkable since
then $\i A(f_{\rm min})$, 
the exponential of which is to be compared with $W_{\rm A}$ 
on the SYM side 
\be
{W_{\rm A}}(\Pi)=\e^{2\i A(f_{\rm min})} \,,
\label{WvsA}
\ee
is real. Here the factor of 2 in the exponent is owing to \eq{A2F}.
Differentiating with respect to $\epsilon$ according to 
Eqs.~\rf{defGamma} and \rf{gnvsGn}, we obtain
\be
\gamma_n=\frac{R^2}{\pi \alpha^\prime} \log{n}
\ee
which reproduces Eqs.~\rf{GKP} and \rf{GKP1} for 
${R^2}/{\alpha^\prime}=\sqrt{\lambda}$.

Some comments concerning the solution~\rf{extremum} are in order.

The inverse function $f(\theta)$ is 
complex-valued for
real $0<\theta<\infty$. Analogously, the $\log^2{L/\epsilon}$-term in
$A(f_{\rm min})$ given by \eq{Sfin} is pure imaginary, while the $A(f)$ itself
is complex in general. 
This property of $A(f_{\rm min})$ agrees with what we had already discussed
at the end of Subsect.~\ref{ss:olambda} for the light-cone Wilson loops in SYM.
There is nothing wrong, in principle, having
such a complex saddle-point trajectory in the path integral.
Its possible quantum-mechanical interpretation will be discussed 
in Subsect.~\ref{ss:qm}.

If we were to use the variables $\r$ and $f$ instead of $\r$ and $\theta$ to
parametrize the worldsheet, then $\theta$ given by \eq{extremum} would be
pure imaginary for $f>1$. Such an imaginary 
\be
\theta=-\i \tau
\label{tau}
\ee 
could be viewed as if $x$ is
pure imaginary, or, in other words, we have performed the analytic 
continuation~\rf{continuing} after which $W$ was expected to become
real. The fact that $A(f)$ is real for $f>1$,
which corresponds to $\tau<\pi(\sqrt{2}-1)/2$,
does not contradict the expected behavior of $W(\Pi)$
in SYM.
When $f$ becomes smaller than $1$
to approach the value $f_{\rm min}$, then
$\tau$ and $A(f)$ become complex. 
One should keep in mind that \eq{HJ} is only an approximation of
the area of the whole minimal surface, spanned by the loop in 
Fig.~\ref{magnify}(a), by its parts near the cusps,
which was used only 
to evaluate the $\log^2(L/\epsilon)$. This minimal surface
is, as has been already mentioned, a nontrivial function of two variables
and is given by the extremum of the action~\rf{NGaction} rather than~\rf{Srt}.

The value $f=1$ is associated with the light cone in $AdS$ space.
Therefore, the values $f<1$ mean that one is no longer inside
the light cone. We discuss an appropriate quantum-mechanical interpretation
of this situation in the next subsection.  


\subsection{Quantum-mechanical interpretation \label{ss:qm}}

After the analytic continuation~\rf{tau} we are dealing
with a saddle-point trajectory of the quantum-mechanical problem
with the time-variable $\tau$, while the right-hand side of \eq{WvsA}
is related to the phase a semiclassical (Schr{\"o}dinger) wave function by
\be
\log{\Psi(f)}  
=\i \frac{2R^2}{\pi \alpha^\prime} \int \frac{\d \r}{\r} \left\{
\int\limits_{f}^\infty \d f\left[ \frac {f^2-1/2}{f\sqrt{f^2-1}}
-1 \right] -f
\right\}.
\label{logPsi}
\ee
We obtain from \eq{extremum}
\be
\tau= - \sqrt{2}\, {\rm arctan}\, \sqrt{2(f^2-1)} 
+{\rm arctan}\,\sqrt{f^2-1}+
\frac {\pi}{2} (\sqrt2 -1 ) \,.
\label{tauextremum}
\ee
The inverse function $f(\tau)$ is now real for 
$\tau\leq{\pi}(\sqrt{2}-1)/{2}$
which means $f\geq 1$. Alternatively, $f$ becomes complex
for $\tau>{\pi}(\sqrt{2}-1)/{2}$. 

As is already mentioned, $A(f)$ is real for $f>1$ and complex
for $f<1$. The wave function in \eq{logPsi} is therefore oscillating
for $f>1$ and exponentially damped for $f<1$.%
\footnote{There could be an additional difference of phases of the wave
function in the two regions, which is calculable either semiclassically
or analyzing the region $f\approx1$.}
 This is a typical behavior for
a quantum-mechanical problem with the potential that forbids for 
a classical particle to penetrate the $f<1$ region. More
arguments in favor of such an interpretation are presented
in Appendix~A, where it is argued that the effective
potential has an infinite wall
at $f=1$: it approaches $-\infty$ when $f$ tends to $1$ from
above and then decreases from $+\infty$. 

The region $f<1$ can be penetrated quantum-mechanically
by tunneling under the barrier, thus
approaching the point $f=f_{\rm min}\approx 1/\sqrt{2}$
that determines the anomalous dimension.
The wave function is then exponentially dumped as is 
displayed in Eqs.~\rf{Sfin} and \rf{logPsi}. 
Note that $\theta_{\rm max}$ cancels on the right-hand side of \eq{logPsi}
which is regular as $f\ra1/\sqrt{2}$.

Once again, this quantum-mechanical interpretation 
of the minimal surface which determines the anomalous dimension
is linked to the mechanical analogy we are considering. 

 
\newsection{Conclusion}

The light-cone Wilson loops are convenient for calculating the
anomalous dimensions of conformal operators of twist two
with large Lorentz spin
both on the SYM and AdS sides.
The lowest-order perturbation-theory calculation in SYM
is in a qualitative agreement with the result given by
the area of the minimal surface formed by the worldsheet of
an open string ending at the loop in the boundary of $AdS$ space.
The latter coincides in turn with the one~\cite{GKP02} obtained
from the classical closed folded string rotating in $AdS_5$.

It would be interesting to pursue the calculations 
in both cases to next orders to compare the results.
There is no problem to calculate the anomalous dimensions
to order $\lambda^2$ in SYM, which is quite similar to the 
calculation of 
Ref.~\cite{KM93} for QCD. The result is to be compared with
an explicit calculation of Ref.~\cite{KL02} to this order.
Analyzing the simplest rainbow graph to the order $\lambda^2$,
we immediately find as in Refs.~\cite{KR87,KM93} that the sum of 
the diagrams with the triple vertex (either of three gauge fields
or one gauge field and two scalars) has to have a term
\be
W^{(2)}\sim \lambda^2 \log^4 \frac{L}{\epsilon}
\ee 
which is required for the exponentiation of the result~\rf{smth}
from the previous order $\lambda$. In contrast to the circular Wilson loop
in Ref.~\cite{ESZ00}, the diagrams with interactions are now present.
Perhaps, they can still be analyzed similarly to Ref.~\cite{DG00}
making a 2D conformal transformation at the boundary, which maps
the straight line~\rf{=1} into a loop of the type in 
Fig.~\ref{magnify}(b). Since it is not a symmetry of $AdS$ 
space, this would rewrite its metric in new coordinates.

The structure of the calculation of the anomalous dimension via 
the minimal surface on the AdS side suggests in turn that it
might be completely determined by a certain state (or states) in 
the (quantum) spectrum of the open string in the AdS background.

\setcounter{section}{0}
\subsection*{Acknowledgments}

I am indebted to Alexander Gorsky, Poul Olesen and Soo-Jong Rey
for very useful discussions.
This work was supported in part by the grant INTAS--00--390.

\appendix{Solving the classical equation}

After the analytic continuation~\rf{tau} the action~\rf{Srt} takes the form
\be
S=\frac{R^2}{\pi \alpha^\prime}\int \frac {\d \r}{\r} 
\int \d \tau \sqrt{\dot f ^2-f^4+f^2 } \,.
\label{Stau}
\ee
Note that the action~\rf{Stau} remains real, so the 
analytic continuation~\rf{tau} is not the same as introducing
an imaginary time for a system with a Hamiltonian which is quadratic in
momentum.

The minimal surface is described by the Euler--Lagrange equation 
\be
\ddot f  V = \dot f^2 V^\prime  -\frac 12 V^\prime V 
\hspace*{2cm}{\rm with~}~V=f^4-f^2\,,
\label{em}
\ee
where $\dot f =\d f / \d \tau $ and 
$V^\prime=\d V/\d f$, which is associated with the 
Lagrangian in \eq{Stau} for $f^4-f^2$ substituted by $V$. 
It is to be solved on the interval $[0,\infty)$ 
with the boundary condition~\rf{b.c.}. 

Equation~\rf{em} is analogous to the equation of motion for
a mechanical system with a velocity-dependent force.
The conserved ``energy'' is given by
\be
E=\frac{V}{\sqrt{\dot f^2 -V }} 
\label{energy}
\ee
which for $E=1/2$ determines 
\be
\dot f = - \sqrt{V+\frac{V^2}{E^2}} =
-\frac{f\sqrt{f^2-1}}{f^2-1/2}
\,.
\label{velocity}
\ee
This reproduces the solution~\rf{tauextremum}.
The ``turning'' point of this trajectory is at $f=1/\sqrt{2}$.
Such a special solution is needed to have a trajectory which gives
the $\log^2{L/\epsilon}$ in \eq{Sfin}.

The value of the momentum along the trajectory is given by
\be
p= \frac{\dot f}{\sqrt{\dot f^2 -V }}=-\sqrt{1+\frac{E^2}{V}}
=-\frac{f^2-1/2}{f\sqrt{f^2-1}}
\label{momentum}
\ee
which results in \eq{logPsi}. We see that $p$ is real
for $f>1$ and becomes infinite as $f\ra1$. Then it is
imaginary for $f<1$. This looks like the effective potential has
an infinite wall at $f=1$: approaching $-\infty$ for $f\ra1$ from
above and then decreasing from $+\infty$. 
Therefore, classical motion is allowed only for $f>1$
(see also Appendix~B).

One can penetrate, however, the region $f<1$ quantum-mechanically
approaching the point $f=f_{\rm min}\approx 1/\sqrt{2}$
by tunneling under the barrier.
The wave function is then exponentially dumped as is 
shown in Eqs.~\rf{Sfin} and \rf{logPsi}.

The region near $f=1$ can be directly analyzed substituting
\be
f(\tau)=1+\xi \e^{g(\tau/\sqrt{\xi})}
\ee
with $\xi\ll1$. Equation~\rf{em} then takes the form
\be
\ddot g = - \e^{-g}
\ee
which coincides with the classical equation of motion for the
Lagrangian
\be
{\cal L}= \frac 12 \dot g^2 +\e^{-g}\,.
\ee
The potential of this problem is $V=-\e^{-g}$ which exponentially
falls down as $g\ra-\infty$, i.e.\ $f\ra1$ from above. 
This is the same assertion as above.

\appendix{Relation to time-like geodesics}

An alternative point of view  on the classical problem of Appendix~A is
as that of constructing geodesics in space with
the metric
\be
\d s^2 = \d f^2 -V(f)\, \d \tau^2 \,.
\label{geometric}
\ee
The equations for time-like geodesics can then be derived from
the Lagrangian which is quadratic in the derivatives of $\tau$ and $f$
with respect
to the proper time $s$ and read as
\bea
\frac{\d}{\d s}\left(V(f) \frac{\d \tau}{\d s} \right) =0\,, 
\label{eqforgeotau}\\
\frac{\d^2 f}{\d s^2} +V^\prime(f)\left(\frac{\d \tau}{\d s} \right)^2=0 
\label{eqforgeof}
\eea
with the constraint
\be
\left(\frac{\d f}{\d s} \right)^2 -V(f) \left(\frac{\d \tau}{\d s} \right)^2 
=1\,. 
\label{constraint}
\ee
Solving the constraint
for $(\d \tau/{\d s})^2$ and substituting into \eq{eqforgeof},
we obtain the equation
\be
V \frac{\d^2 f}{\d s^2} + V^\prime 
\left[\left(\frac{\d f}{\d s} \right)^2-1\right] = 0
\label{fgeo}
\ee
which determines the geodesic $f(s)$.

A convenient way to solve \eq{fgeo} is as follows.
From \eq{eqforgeotau} we have
\be
\frac{\d \tau}{\d s} = \frac{E}{V }\,,
\ee
where $E$ is an integration constant. Substituting in \eq{constraint},
we find
\be
\frac{\d f}{\d s} = - \sqrt{1+\frac{E^2}{V}}
\label{thesame2}
\ee
which yields
\be
\frac{\d f}{\d \tau} = -\sqrt{V+\frac{V^2}{E^2}}\,.
\label{thesame}
\ee
Equations~\rf{thesame2} and \rf{thesame} are the same as Eqs.~\rf{momentum}
and \rf{velocity} 
for $V=f^4-f^2$ 
and  $E=1/2$. The conclusion is the same:
these  geodesics also exist
only for $f>1$ where the signature of
the metric~\rf{geometric} is Lorentzian.


\end{document}